# Status of the LBNE Neutrino Beamline

**Vaia Papadimitriou**

L2 Manager for the LBNE Neutrino Beamline,
Fermilab, PO Box 500, Batavia, Illinois, 60510, U.S.A

vaia@fnal.gov

**Abstract**. The Long Baseline Neutrino Experiment (LBNE) will utilize a neutrino beamline facility located at Fermilab to carry out a compelling research program in neutrino physics. The facility will aim a beam of neutrinos toward a detector placed at the Homestake Mine in South Dakota. The neutrinos are produced in a three-step process. First, protons from the Main Injector (60-120 GeV) hit a solid target and produce mesons. Then, the charged mesons are focused by a set of focusing horns into the decay pipe, towards the far detector. Finally, the mesons that enter the decay pipe decay into neutrinos. The parameters of the facility were determined taking into account several factors including the physics goals, the Monte Carlo modeling of the facility, spacial and radiological constraints and the experience gained by operating the NuMI facility at Fermilab. The initial beam power is expected to be ~700 kW, however some of the parameters were chosen to be able to deal with a beam power of 2.3 MW. We discuss here the status of the conceptual design and the associated challenges.

### 1. Introduction

The Neutrino Beamline is the central component of LBNE, and its driving physics considerations are the long baseline neutrino oscillation analyses. These include the search for, and precision measurement of, the parameters that govern $\nu_\mu$ to $\nu_e$ oscillations as well as precision measurements of $\theta_{23}$ and $|\Delta m^2_{32}|$ in the $\nu_\mu$ disappearance channel. On January 8, 2010, the Department of Energy approved the "Mission Need" for LBNE and the LBNE Project is developing a conceptual design to meet this Mission Need.

The Neutrino Beamline facility is expected to be fully contained within Fermilab property. The primary proton beam, in the energy range of 60-120 GeV, will be extracted from the Main Injector [1] using "single-turn" extraction. For 120 GeV operation and with the Main Injector upgrades implemented for the NOvA experiment [2], the fast, single turn extraction will deliver all the protons ($4.9 \times 10^{13}$) in one machine cycle (1.33 sec) to the LBNE target in 10 μs. The beam power is expected to be ~700 kW in the energy range of 80 to 120 GeV. The charged mesons produced by the interaction of the protons with the target are focused by two magnetic horns into the decay pipe towards the far detector. These mesons are short-lived and decay into muons and neutrinos. At the end of the decay region, an absorber pile is needed to remove the residual particles remaining at the end of the decay pipe. The neutrino beam is aimed at the Homestake Mine in South Dakota, about 1300 km away, with a $5.8^o$ vertical bent.

A wide band neutrino beam is needed to cover the first and second neutrino oscillation maxima, which for a 1300 km baseline are expected to be approximately at 2.3 and 0.8 GeV. The beam must provide a concentrated neutrino flux at the energies bounded by the oscillation peaks and we are therefore optimizing the beamline design for neutrino energies in the area 0.5 to 5 GeV. The initial operation of the facility will be at a beam power incident on the production target of ~700 kW, however some of the initial implementation will have to be done in such a manner that operation at 2.3 MW can be achieved without retrofitting. Such a higher beam power is expected to become available in the future with the planned Project X [3] which includes the replacement of the existing proton source that feeds the Main Injector. In general, components of the LBNE beamline system which cannot be replaced or easily modified after substantial irradiation at 700 kW operation are being designed for 2.3 MW. Examples of such components are the shielding of the target chase and the decay pipe, and the absorber.

The LBNE Neutrino Beamline design has to implement as well stringent limits on the radiological protection of the environment, workers and members of the public. The relevant radiological concerns, prompt dose, residual dose, air activation and tritium production have been extensively modeled and the results implemented in the system design. A software model for the entire beamline has been coded into the MARS15 simulation package [4]. A most important aspect of modeling at the present design stage is the determination of the necessary shielding thickness and composition in order to protect the ground water and the public and to control air emissions.

This paper is a snapshot of the present status of the design as exemplified in the current Conceptual Design Report [5,6].

## 2. The Neutrino Beamline Reference Designs

Between January, 2009 and September, 2010 the LBNE Neutrino Beamline team developed and costed a NuMI-style Neutrino Beamline conceptual design with the facility located mostly in rock. From October, 2010 and on we entered the $2^{nd}$ and $3^{rd}$ phase of value engineering effort with the goal to reduce significantly the cost of the entire LBNE project. The Neutrino Beamline team developed and evaluated 17 value engineering proposals and defined and costed four beamline configurations with varying primary beam extraction points and beamline depth. For two of the configurations we considered a longer beamline extracting from the MI-60 straight section of the Main Injector (same as for NuMI) and for two of them considered a shorter beamline extracting from the MI-10 section. For each extraction point we considered a deep option with excavation in soil and in rock and a shallow option featuring a large berm into which facilities would be constructed.

In the end of June, 2011 we selected two of the four beamline configurations to develop further towards the conceptual design review, and we applied on them as many, smaller scope, value engineering proposals as possible. These two reference designs are being refered to as MI-60, Deep and MI-10, Shallow. They both feature a 200-250 m long decay pipe with a diameter of 4 m and a muon range-out distance between the absorber and the Near Detector of 210 m.

### 2.1. The MI-60, Deep beamline configuration

Figure 1 shows a longitudinal section of the Neutrino Beamline facility in the MI-60, Deep beamline configuration. In this case the beam extraction enclosure exists and is being shared with NuMI/NOvA. A dipole magnetic beam switch is supposed to take the beam into either the LBNE or NOvA beam transport channels. A group of technical components bend the proton beam $48^o$ westward and $5.8^o$ downward to establish the final trajectory toward the far detector. There is sufficient space available within Fermilab property in this configuration to increase significantly, if needed, either the decay pipe length or the muon range-out distance.

The radiological model in the deep option is such that groundwater is encouraged to migrate through the rock mass toward and into the decay region where it can be collected and transported away. The decay pipe shielding thickness has been studied as a function of rock porocity and water inflow and has been determined to be 3.3 m of concrete between the pipe and the native rock.

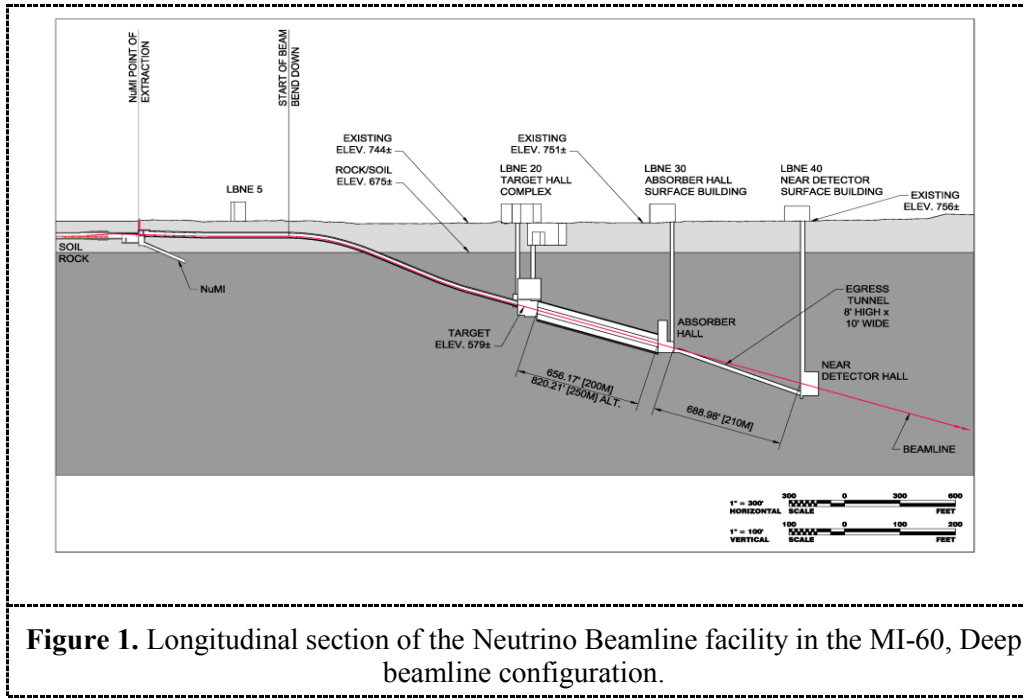

**Figure 1.** Longitudinal section of the Neutrino Beamline facility in the MI-60, Deep beamline configuration.

2.2. The MI-10, Shallow beamline configuration
Figure 2 shows a longitudinal section of the Neutrino Beamline facility in the MI-10, Shallow beamline configuration. At MI-10 there is no existing extraction enclosure and we are minimizing the impact on the Main Injector by introducing a 15.6 m long beam carrier pipe to transport the beam through the Main Injector tunnel wall into the new LBNE enclosure. A group of technical components send the proton beam through an above-grade beamline covered by a man-made embankment/hill whose apex is ~22 m and its footprint ~318,000 sq. ft. The beam then will be bent downward toward a target located at grade level. The overall bend of the proton beam is $7.2^o$ westward and $5.8^o$ downward to establish the final trajectory toward the far detector. One of the major differences with the MI-60, Deep option is that here the primary beam enclosure and the Target Hall complex are founded on rock using drilled shafts.

In the shallow option, because of the presence of a local aquifer at and near the top of the rock surface we cannot encourage groundwater to migrate toward and into the decay region (significant daily collection) and therefore we have to provide an engineered geomembrane barrier between the shielding and the environment. The decay pipe shielding thickness has been determined to be 5.5 m of concrete (see Fig. 3). In this configuration, since the Target Hall and Decay Pipe are not deep into rock, we also had to check and verified that the total radiation dose at the Fermilab boundary is about four orders of magnitude below the LBNE related allowed dose of 1mrem/year .

**3. Status of the conceptual design**
The LBNE Neutrino Beamline scope inludes a Primary Beam, a Neutrino Beam and a few "Integration Systems" related to both Primary and Neutrino Beam, like Alignment, Controls,

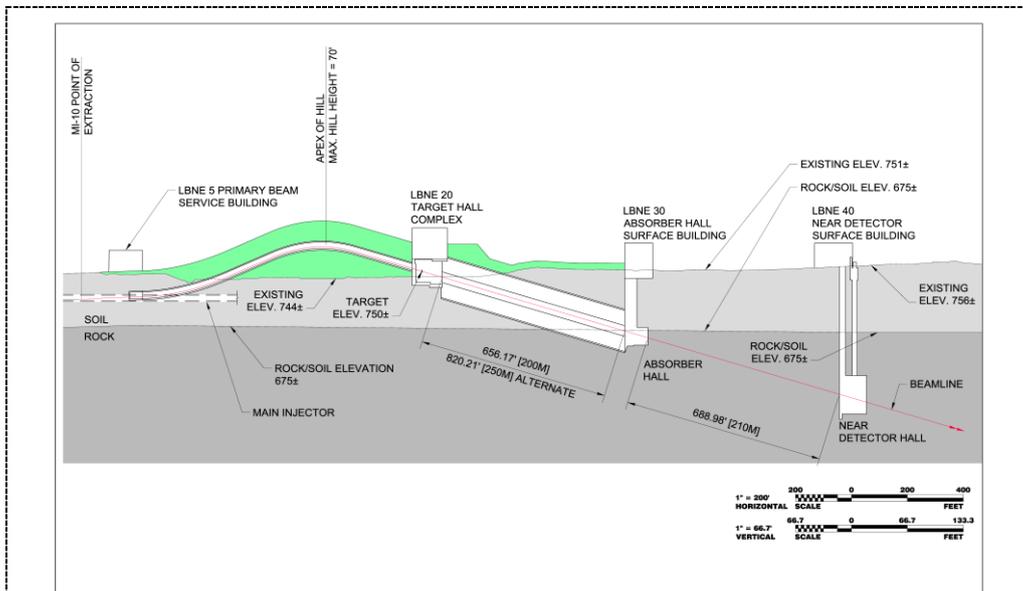

**Figure 2.** Longitudinal section of the Neutrino Beamline facility in the MI-10, Shallow beamline configuration.

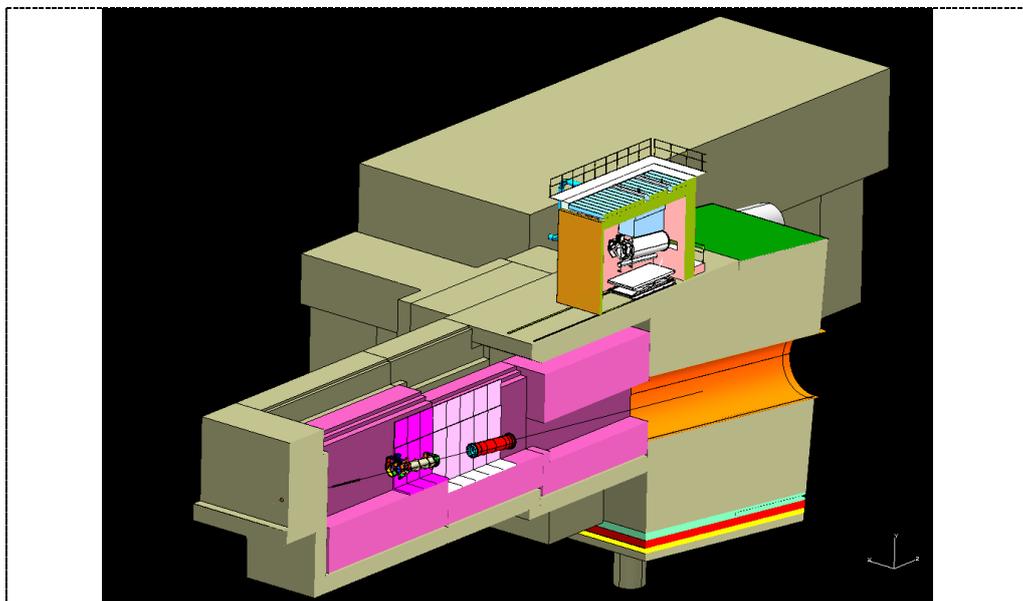

**Figure 3.** Section of the Target Hall complex and the Decay Pipe in the MI-10, Shallow configuration.

Interlocks, and Installation. The Primary Beam elements necessary for transport include vacuum pipes, dipole, quadrupole and corrector magnets and beam monitoring equipment (Beam Position Monitors, Beam Loss Monitors, Beam Profile Monitors, etc.). The magnets are conventional, Main Injector design, and the magnet power supplies are a mixture of new, Main Injector design power supplies or refurbished Tevatron power supplies. The beam optics accommodates a range of spot sizes on the target in the energy range of interest and for beam power up to 2.3 MW, and the beam transport is expected to take place with minimal losses. All of the conceptual design effort for the Primary Beam is taking place at Fermilab.

The Neutrino Beam includes in order of placement (1) a baffle collimator assembly to protect the target and the horns, (2) a target, (3) two magnetic horns. They are all located inside a heavily shielded vault, called the target chase, that is open to the decay pipe at the downstream end. An air-filled, air-cooled decay pipe follows which allows pions and kaons to decay to neutrinos, and in the end of it is the absorber which is a specially designed pile of aluminum, steel and concrete blocks, some of them water cooled, which must contain the energy of the particles that exit the decay pipe. Radiation damage and cooling of elements, tritium mitigation, remote handling and storage of radioactive components are essential considerations for the conceptual design of the Neutrino Beam. Most of the design effort for the Neutrino Beam is taking place at Fermilab except from the designs of the primary beam window, the target, the horn support modules and the remote handling for which we have established collaborations with other institutions.

The reference design of the target assembly is based on work done at the Institute of High Energy Physics (IHEP) in Protvino, Russia [7]. The target is a 95 cm long cylinder consisting of 48 small cylinders with diameter of 15.3 mm and spaced 0.2 mm apart. The target material is POCO ZXF-50 graphite and it is water cooled. In order to capture low energy pions and kaons from the target at large production angles, the target assembly is inserted well into the first horn's magnetic lens whose inner conductor is cylindrical/parabolic in shape. We are in the process of comparing the properties and longevity of the POCO graphite with other target sample materials which we have irradiated at the BLIP facility of the Brookhaven National Laboratory. Another approach is to consider beryllium as an alternative target which is expected to be more resistant to radiation than graphite. Colleagues from Rutherford Appleton Laboratory (RAL) have analyzed the issues associated with a potential beryllium target and have concluded that a 13 mm diameter, 1 m long cylinder (or a segmented target) falls inside the chosen design point stress for 700 kW operation [8].

The technical part of the conceptual design effort for the Neutrino Beamline (including the "Integration Systems") is almost complete at this point. The Neutrino Beamline team is in the process of implementing improvements in the design while refining costs and schedules.

### 4. Conclusion

The LBNE Neutrino Beamline has been at a technical status suitable for a conceptual design review for a NuMI-style beamline, mostly in rock, since September, 2010. Since then we developed and reviewed several value engineering proposals with the goal of reducing the cost further. We have considered and costed four beamline configurations at different depths and primary beam extraction points from Fermilab's Main Injector, and in the end of June, 2011 we selected two of them as reference designs to be developed further. We developed conceptual design reports and a set of risks for both of these designs.

After a thorough review of both conceptual designs, on November 10, 2011, we selected as the default beamline configuration the MI-10, Shallow option. We plan to develop this beamline concept fully, aiming for an LBNE conceptual design review in the Spring of 2012. The baseline review is anticipated for the summer of 2013.